\documentclass[12pt,journal]{IEEEtran}
\usepackage[latin1]{inputenc}
\usepackage[T1]{fontenc}
\ifCLASSOPTIONcompsoc
  \usepackage[nocompress]{cite}
\else
  \usepackage{cite}
\fi

\usepackage{amsmath}
\usepackage[pdftex]{graphicx}

\usepackage{url}

\newcommand{\wabs}[1]{\left|#1\right|}
\newcommand{\wcal}[1]{\mathcal{#1}}
\newcommand{\wfc}[2]{{#1}\!\left(#2\right)}
\newcommand{\wref}[1]{(\ref{#1})}
\newcommand{\wvec}[1]{\mathbf{#1}}

\begin{document}
\title{Moore: Interval Arithmetic in Modern C++}
\author{W.~F.~Mascarenhas
\IEEEcompsocitemizethanks{
\IEEEcompsocthanksitem Walter Mascarenhas works at
the Institute
of Mathematics and Statistics, Universidade~de~S\~{a}o~Paulo, Brazil.
His e-mail address is
 walter.mascarenhas@gmail.com.
}}
\markboth{Submitted to IEEE Transactions on Computers}
{Moore: Interval Arithmetic in modern C++}

\IEEEtitleabstractindextext{%

\begin{IEEEkeywords}
Interval Arithmetic, C++, Concepts.
\end{IEEEkeywords}}

\maketitle

\IEEEpeerreviewmaketitle

\ifCLASSOPTIONcompsoc
\IEEEraisesectionheading{\section{Introduction}\label{sec:introduction}}
\else
\section{Introduction}
\label{sec:introduction}
\fi

\IEEEPARstart{I}{nterval} arithmetic is a powerful tool,
which can be used to solve practical and theoretical
problems. However, implementing an interval arithmetic library
for general use is difficult, and requires much attention to detail.
These details involve the specification of interfaces,
the implementation of efficient and accurate algorithms
and dealing with bugs, which can be our own fault or
the compiler's.

Regarding interfaces, there are many design decisions
to be made and each author has his own opinions,
as one can see by comparing the libraries in
\cite{Boost:boost,CXSC:cxsc, Filib:filib,
Gaol:gaol,Gnu:gnu,JInterval:jinterval,Lambov:lambov,
MPFI:mpfi,NehmeierA:nehmeierA,Profil:profil,Sun:sun}.
Recently two documents \cite{P17881:full,P17881:p17881}
were elaborated in order to provide a
common interface for interval arithmetic libraries.
Formally, the first document \cite{P17881:full} is an active
IEEE standard, while the second \cite{P17881:p17881} is still a project.
For simplicity, throughout the article we will refer to
these documents as ``The IEEE standards.''
These standards are a significant step forward,
but there is controversy in the interval arithmetic
community regarding them. For instance,
prof. U. Kulisch has made clear his disagreement
with them.

This article presents the Moore library, which
implements part of the  IEEE
standards  in the most recent version of the C++ language,
using new features of this language. The library was written
mostly to be used on our own research, and we
focus on improving the code more likely to be used often.
For this reason, we do not fully implement the IEEE standards.
However, we do hope that the library will be useful
to others. People for which compliance with the standards
is a priority would be better served by the libraries in
\cite{Gnu:gnu} and \cite{NehmeierA:nehmeierA};
people looking for better performance or
more precise types of endpoints for their intervals
may consider using our library. For instance,
in Section \ref{sec:experiments} we present
experiments showing that our library is competitive
in terms of speed with well known libraries, and it
is significantly faster than the current reference implementation
of the IEEE Standards in C++ \cite{NehmeierA:nehmeierA}
(the other reference implementation, \cite{Gnu:gnu}, is implemented
in Octave and is not comparable to our library.)

This article was written for people which are
already familiar with interval arithmetic, who
will understand us when
we say that, when used properly, our library satisfies
all the usual containment requirements of interval arithmetic.
Our purpose is to describe the library,
show that it is competitive with well known libraries,
and expose its limitations (see the last section.)
Even with these limitations, we would like to invite
readers to experiment with the library. In fact,
by reading this article one will have only
a glimpse of the library, and the only way
too fully understand it is to try it in practice.
The Moore library is open source
software, distributed under the Mozilla 2.0 license,
and its source code can be obtained by
sending an e-mail message to the author.

In the rest of this article we present
the library, starting from the basic arithmetic
operations and moving to more advanced issues.
We also describe
in which points our library deviates from the IEEE standards
and the parts of the standards that we do implement.

\section{Hello Interval World}
\label{sec:hello}
The Moore library can be used by people
with varying degrees of expertise. Non
experts can simply follow what is outlined
in the code below:
{\small
\begin{verbatim}
#include "moore/minimal.h"

Moore::RaiiRounding r;
Moore::Interval<> x(2.0, 3.0);
Moore::Interval<> y("[-1,2]");

for(int i = 0; i < 10; ++i)
{
  y = (sin(x) - (y/x + 5.0) * y) * 0.05;
  cout << y << endl;
}
\end{verbatim}
}
\noindent
In English, with the Moore library
one can construct intervals by providing their endpoints
as numbers or strings, and then use
them in arithmetical
expressions as if they were numbers. The
library also provides the trigonometric and
hyperbolic functions, their inverses,
exponentials and logarithms,
and convenient ways to read and write
intervals to streams.

The file {\small \verb minimal.h } in the code above
contains the required declarations for
using the library with {\small \verb double } endpoints.
The line
{\small \
\begin{verbatim}
Moore::RaiiRounding r;
\end{verbatim}
}
\noindent
is required in order to use the library. It sets the rounding mode to upwards,
and the rounding mode is restored
when {\small \verb r } is destroyed,
following the Resource Acquisition is Initialization (Raii)
pattern in C++. Our approach is similar to one
option provided by the boost library \cite{Boost:boost}.
However, the boost library is more
flexible than ours: we only provide one rounding
policy. In fact, providing fewer options instead
of more is the usual choice
made by the Moore library.
We only care about concrete use cases motivated by
our own research, instead of all possible uses of interval arithmetic.
This is the main difference between the spirit of
our library and the purpose of the
IEEE standards for interval arithmetic.
We prefer to provide a better
library for a few users rather than
trying to please a larger audience which
we will never reach.

The intervals in the Moore library are parameterized
by a single type. It does not contain class hierarchies,
virtual methods or policy classes.
On the one hand, users can only choose
the type of the endpoints defining the intervals
of the form $[a,b]$ with $-\infty \leq a \leq b \leq +\infty$,
or the empty interval.
On the other hand, we do believe that our library goes
beyond what is offered by other libraries
in its support of generic endpoints and operations. As
we explain in Section
\ref{sec:ends},  the library can
work with several types of endpoints
``out of the box,'' that is, it
provides tested code in which several
types of endpoints can be combined,
as in this example:
{\small
\begin{verbatim}
RaiiRounding r;
Interval<>           x("[-1,2]");
Interval<float>      y("[-1/3,2/3]");
Interval<__float128> z("[-inf,4"]);
Interval<Real<256>>  w("5?");

auto h = hull(x, y, 0.3);
auto i = intersection(x, y, z, w);
auto j = sin(z * x/cos(y * z)) - exp(w);
\end{verbatim}
}
\noindent
The code above handles four kinds of endpoints:
\begin{itemize}
\item The interval {\small \verb x } has endpoints of type {\small \verb double }.
\item {\small \verb y } has endpoints of  type {\small \verb float }.
\item The endpoints of {\small \verb z } have quadruple precision.
\item {\small \verb w } has endpoints of type {\small \verb Real<256> }, which
represents floating point numbers with $N=256$ bits of mantissa,
and the user can choose other values for $N$.
\item The compiler deduces that {\small \verb i } is an interval with endpoints of type
{\small \verb double }, which is the appropriate type for storing the convex hull of
{\small \verb x }, {\small \verb y } and {\small \verb 0.3 }.
\item It also deduces that {\small \verb Real<256> } is the appropriate type of endpoints
for the intervals representing the intersection of {\small \verb x, y, z } and
{\small \verb w } and the result of the evaluation of the expression assigned to {\small \verb j }.
\end{itemize}

We ask the reader not to underestimate
the code in the previous paragraph. It
is difficult to develop the
infrastructure required for users to
handle intervals with endpoints of
different types in expressions as natural
as the ones in that code.
In fact, there are numerous
issues involved in dealing with intervals
with generic endpoints, and simply writing
generic code with this purpose is not enough.
The code must be tested, and our experience
shows that it may compile for some types
of endpoints and may not compile for others.

There are two main points in which the Moore
library does not follow the IEEE standards:
decorations and exceptions. As we
explain in Section \ref{sec:dec},
we do not provide decorated intervals, because
in our opinion decorations are a bad idea.
We agree that it is quite useful to have
standards, and we are glad to acknowledge
the positive influence of the IEEE standards
in our library. However, we also believe
that standards should cover only the minimal number
of features required to achieve the goals
of the majority of the users, and
decorations do not satisfy this basic
criterium. We should also mention that we
did implement a library which supported
decorations, and that when we
analyzed the result it was evident that
this implementation should be thrown away.

We also treat exceptions differently. In the Moore
library, the term exception applies to an error from
which no recovery is possible (or desirable), and such
that its occurrence leads to the termination of the program.
For instance, as the excellent GMP library, we find it
appropriate to terminate the program when we are not
able to allocate memory. In fact, any library relying
on GMP would be subject to program termination in case
of errors in memory allocation, unless precautions are taken.
The Moore library uses GMP and did not take any precautions.

In our library, warnings are indicated by return values instead of raising flags.
For instance, instead of raising an inexact flag when constructing
an interval from the string {\small \verb "[1/3,2/3]" } we
provide a constructor with a second argument of type
{\small \verb Accuracy } which indicates the accuracy with which
the string is converted to an interval, as in
{\small
\begin{verbatim}
Accuracy a;
Interval<> x("[1,2]", a);
Interval<> y("[1/3,2/3]", a);
Interval<> z("[1,2]");
Interval<> w("[1/3,2/3]");
Interval<> bad("disaster");
\end{verbatim}
}
In the construction of {\small \verb x },
the variable {\small \verb a } will be set to
{\small \verb Accuracy::Exact }, and the construction
of {\small \verb y } yields {\small \verb  a  \verb =  \verb Accuracy::Tight }.
In the other constructor calls the user is not interested in knowing
the accuracy and there is no reason to set a flag to
inform him about something that he does not care.
In debug mode, the construction of {\small \verb bad } will
cause the failure of an assertion, and the program will
be stopped by the debugger. In release mode, the construction
of {\small \verb bad } will terminate the program, when
calling an error handler which writes a
message to the standard output stream for errors.
This the default behavior, and the user can change it
by changing the error handler, but we do not
recommend it.  We discuss Exception handling in more detail
in Section \ref{sec:except}.

Exceptions in the Moore library are reserved for truly exceptional cases,
and when in doubt whether a string represents an interval the user
should do something like the code below instead of trying to construct
the interval directly from it:
{\small
\begin{verbatim}
Accuracy a;
Interval<> x;
a = try_parse(x, "Am I an interval?");

if( a == Accuracy::Invalid ) {
  cout << "No, You are not an interval";
}
\end{verbatim}
}

Finally, other than the cases mentioned above,
one case mentioned in the next section,
and the implementation of the functions
{\small \verb cancel_minus } and
{\small \verb cancel_plus },
the Moore library follows the IEEE standards
for interval arithmetic closely, because they
contain many good points.
In particular, we implement all the functions mentioned in the
simplified standard \cite{P17881:p17881}, including
the ones which would qualify our library as
a flavor of the full standard \cite{P17881:full} in the absence
of the points mentioned in the previous paragraph.
We also implement the reverse and overlapping
functions mandated by the full standard \cite{P17881:full}.
In the end, people not interested in decorations and
exceptions will be able to use our library as they
would use any other library conforming with the
standards, with the additional functionality
described in the next sections.

\section{Input and output}
\label{sec:io}
Except for minor details which are still under
discussion in working group for the simplified
standard \cite{P17881:p17881}, we agree with
all points mandated by the IEEE standards regarding
input and output of plain intervals, and tried to
implement them as faithfully as we could. We
also provide several options for the formatted
output of intervals, as an aid to the
visual inspection of the results of interval calculations.

For instance, the code in the first page of this article
outputs ten lines, and the first one is
{\small
\begin{verbatim}
[-0.592944,0.345465]
\end{verbatim}
}
\noindent
Users may want to display this interval with more digits, or to
use scientific notation. With the Moore library
they could write
{\small
\begin{verbatim}
cout << std::setprecision(10);
cout << std::scientific;
\end{verbatim}
}
\noindent
before the {\small \verb for } statement in that code, and obtain
{\small
\begin{verbatim}
[-5.929439996e-1,3.4546487135e-1]
[-8.2293866866e-2,1.9882190431e-1]
...
\end{verbatim}
}
They could also introduce space between
the numbers and the brackets and pad the numbers
with zeros to the right, by including the lines
{\small
\begin{verbatim}
cout << Moore::Io::pad();
cout << Moore::Io::border_slack(2);
cout << Moore::Io::center_slack(1);
\end{verbatim}
}
\noindent
which lead to the better looking output
{\small
\begin{verbatim}
[  -5.9294399960e-1, 3.4546487135e-1  ]
[  -8.2293866866e-2, 1.9882190431e-1  ]
...
\end{verbatim}
}

There are several options for
controlling the way in which intervals
are written to output streams, and
they can also be read from input
streams as in
{\small
\begin{verbatim}
Moore::Interval<> x;
cin >> x;
Accuracy a = try_scan(x, is);
\end{verbatim}
}

In order to simplify its use, most options in the
Moore library have reasonable default values, which
are not always the most efficient. For instance,
the convenience of io streams has a cost. Usually
this overhead does not matter, but when reading
and writing large files we can improve
things a bit as in the next example, which
reads intervals from a file and writes
them to another file

{\small
\begin{verbatim}
  std::vector<Interval<>> in, out;
  ....
  std::ofstream os("file.txt");
  write(out, os, Io::Format("A"));
  os.close();

  std::ifstream is("file.txt");
  read(in, is);
  is.close();

  std::ifstream is2("file.txt");
  Accuracy a = try_read(in, is);

  assert( a == Accuracy::Exact );
\end{verbatim}
}
In the code above we use the {\small \verb Format }
object in order to write the interval
in hexadecimal format (as with "\%A" in {\small \verb printf }),
and ignore the flags of the ostream os.
This ensures that intervals will be read exactly
as they were written. The example in Section \ref{sec:hello}
is more inefficient, because for each call of
{\small \verb operator<< } we would read the flags of
{\small \verb std::cout }, create a corresponding
{\small \verb Format } object and then
write the interval according to this format, using dynamic
allocated memory. In the example above there is
only one {\small \verb Format  } object and only a few dynamic memory
allocations.

\section{Endpoints}
\label{sec:ends}
There is only one interval class in the
Moore library, and it
has a  single parameter: the type of the endpoints.
The library provides a few types of endpoints
``out of the box'', that is, endpoints which
are ready to be used and have been tested. The library
was designed, implemented and tested in order to
provide the speed of arithmetics supported by
the hardware, with the types
{\small \verb float }  and {\small \verb double },
efficient quadruple precision,
with the type {\small \verb __float128 }, and
high precision based on the MPFR library
\cite{MPFR:mpfr}. It also allows for
operations mixing different types of
endpoints. For instance, we could use
a less precise and more efficient type
for the $x$ coordinate and a more precise
type of endpoint for the $y$ coordinate
or in intermediate computations.

In summary, we currently allow the combination of following
types of endpoints:
\begin{itemize}
\item The standard floating point types:
{\small \verb float },
{\small \verb double } and
{\small \verb long  \verb double }.
\item The quadruple precision type
{\small \verb __float128 } provided by
the gcc's quadmath library.
\item Floating point numbers with
mantissa of $N$ bits, where $N$ is
a constant fixed at compile time. We
provide a type {\small \verb Real<N> },
which is a stack based wrapper of the
{\small \verb __mpfr_struct } from the
MPFR library \cite{MPFR:mpfr}.
\end{itemize}

\section{Decorations}
\label{sec:dec}
A significant part of the IEEE standards for interval arithmetic
is devoted to {\it decorations}. Basically, a decoration is
a tag attached to an interval in order to provide
information regarding how it was obtained. The combination
of the interval and the tag is called by {\it decorated interval},
and the standards have several requirements regarding
decorated intervals.

In principle, decorated intervals would be a convenient
way to propagate information about the evaluation of
functions and exceptions in a thread safe way.
However, we do not plan to use decorations, or
to support them in the Moore library, because we
believe that the cost involved in implementing,
testing and maintaining the resulting code outweighs the
benefits that decorations may bring, specially
when we consider the use of various types of
endpoints.

The problems with decorations start already
in the simple case of double precision endpoints.
With g++, a struct like
{\small
\begin{verbatim}
struct Interval {
  double inf;
  double sup;
};
\end{verbatim}
}
\noindent
uses 16 bytes of memory, and adding a
tag of type {\small \verb char } leads to
a decorated interval of type
{\small
\begin{verbatim}
struct DecInterval {
  double inf;
  double sup;
  char decoration;
};
\end{verbatim}
}
\noindent
which occupies 24 bytes when compiled with
g++, and this is 50\% more than
the memory used by plain intervals. Moreover, tags
affect the alignment of the resulting struct:
Today's cache lines usually have 64 bytes,
and  can hold four {\small \verb Intervals },
but only two {\small \verb DecIntervals. } Therefore,
cache misses will happen more often when using
decorated intervals, and access to memory
will be more expensive, specially
for vectors and matrices.

Another important issue is the combinatorial
explosion of cases to be considered while
planing, coding and testing libraries combining
plain and decorated intervals. Each function
with two arguments would require
a version for each one the four possible
kinds of pairs of inputs.
Moreover, we would need
to define the resulting decoration for all
possible combinations of the decorations of
the inputs, and there is no universal rule
for defining the decoration of the output.
For example, the functions
{\small \verb convexHull } and {\small \verb intersection }
in the IEEE standards have the decorations
of the output defined in an ad-hoc way,
and the users of the library will need to
think about the decoration of the output
of every function they write in order to
keep the overall consistency of the decoration
system. Finally, functions like
fma and some reverse functions, which take three arguments,
would require eight versions, and each one of them
would need to be planed, coded and tested.

As a result of this combinatorial explosion on the
code size caused by decorations, one of the two
reference implementations
of the IEEE standards \cite{NehmeierA:nehmeierA}
does not implement functions combining plain
and decorated intervals with endpoints of different
types. The other reference implementation
\cite{Gnu:gnu} handles only endpoints of type
{\small \verb double } and is not as affected by
the large number of possibilities entailed by
the combination of plain interval, decorated
interval and different types of endpoints.
Therefore, we are not alone in our need
to make strategic decisions regarding
what should and should not be implemented.
We decided to prioritize generic endpoints,
the author of \cite{Gnu:gnu} chose
to implement only endpoints of type {\small \verb double },
and the author of \cite{NehmeierA:nehmeierA} did as much
as he found reasonable in order to support
both decorations and generic endpoints.
He also chose not to implement operations involving numbers
and intervals, or the accumulation operators +=,-=,*= and /=
for intervals.
In the end, we believe that there is a place for
each one of us among the users of
interval arithmetic libraries.

\section{Exceptions}
\label{sec:except}
The Moore library handles exceptions depending
on the mode in which the code is compiled. There
are three modes: Debug, Fast and Safe.

Debug mode is in effect when {\small \verb NDEBUG }
is not defined. In this case {\small \verb asserts }
check the input to functions and whether the
rounding mode is upwards (with other rounding modes the
Moore library violates the usual rules of interval arithmetic.)
In this mode the debugger stops
the program when an inconsistency is found and the user will
know in which point of the code the problem is.
The downside of these safeguards is their cost.
For instance, if statements and calls to
{\small \verb std::fegetround } in inner loops can
have a noticeable negative effect. Debug
mode is the safest one for the library and we suggest
it for less experienced users.

The Fast mode is used when {\small \verb NDEBUG } is defined
and {\small \verb MOORE_IN_SAFE_MODE } is not defined. It is
appropriate for users secure about the
correctness of their code, because it performs
little checking, in cases in which
we expect the overhead to be minimal. In this
mode, if an error is detected then an error handler is called.
The default error handler is a function {\small \verb on_error },
declared in the file {\small \verb exception.h }.
The file {\small \verb on_error.cc } contains
a definition of this function, which writes a message to the standard
output stream for errors and terminates the program.
The user can replace this error handler by linking his
own function {\small \verb on_error }.

The Safe mode is used when both {\small \verb NDEBUG } and {\small \verb MOORE_IN_SAFE_MODE }
are defined. In this mode we check for the same issues considered in
Debug mode, and pay the same overhead, but instead of using
assertions we call the error handler mentioned in the previous paragraph
in case an inconsistency is detected. This mode is useful when one is trying to
find bugs introduced by the optimizer. Dealing with
this kind of bug is difficult because they disappear in Debug builds, but
they are a part of life and we provide the Safe mode to help users
and ourselves to deal with them.

\section{Experiments}
\label{sec:experiments}
In this section we present the results of experiments comparing
the Moore library with three other interval arithmetic libraries:
Boost Interval \cite{Boost:Bst}, Filib \cite{Filib:filib}
and libieeep1788 \cite{NehmeierA:nehmeierA}. In summary,
we show that our library is slightly faster than the Boost library,
it is significantly faster than the libieeep1788 library, and
it is faster than the Filib library in applications which rely
only on arithmetic operations, but the elementary functions
(sin, cos, etc.) in Filib are significantly faster than our
library, the boost library and the libieeep1788 library.
This difference in the evaluation of the elementary functions
happens because the Boost, Moore and libieeep1788
libraries use the MPFR library, whereas Filib has its own
implementation of the elementary functions, which
was optimized for IEEE 754 double precision.

\begin{table}[!t]
\renewcommand{\arraystretch}{1.3}
\caption{Normalized Times for the Lebesgue Function}
\label{table:lebesgue}
\centering
\begin{tabular}{cccc}
\hline
\texttt{Moore} & \texttt{Filib} & \texttt{Boost} & \texttt{P1788}\\
1     & 3.8   & 1.1   & 268.5 \\
\hline
\end{tabular}
\end{table}

Besides the difference in speed, there is a difference in
the accuracy of the elementary functions. When using
IEEE754 double precision, due to the way
in which argument reduction is performed, the Boost
and Filib libraries can lead to errors of
order of the square root of the machine precision ($10^{-8}$)
in situations in which the Moore library and the libieeep1788 library
lead to errors of the order of the machine precision
$(10^{-16})$. Therefore, there is a trade off between accuracy
and speed between the Moore and the Filib libraries.
The accuracy of the elementary functions provided
by the Filib library suffices for many applications
and it would be a better choice than the Moore library
for application in which the use of such functions
with accuracy of order $\wfc{O}{\sqrt{\epsilon}}$
would suffice, and such functions would be called
often.

The Moore library was implemented to be used
in our research, and the experiments presented
in this section reflect this. We present timings
for situations related to our current research about
the stability of barycentric interpolation \cite{MascA,MascB,MascC}.
In this research we look for parameters $w_0,\dots w_n$ which
minimize the maximum of the {\it Lebesgue function}
\begin{equation}
\label{lebesgue}
\wfc{\wcal{L}}{\wvec{w};\wvec{x},t} :=
\ \ \left.
\sum\limits_{k = 0}^{n} \wabs{\frac{w_k}{t - x_k}} \right/ \wabs{\sum\limits_{k = 0}^{n} \frac{w_k}{t - x_k}}
\end{equation}
among all $t \in [-1,1]$, for a given vector $\wvec{x}$ of nodes,
and we use interval arithmetic to find such minimizers and
validate them.

In our first experiment we measured the time to evaluate the
Lebesgue function in \wref{lebesgue} for 257 Chebyshev nodes
of the second kind \cite{MascA}, with interval eights,
at a million points $t$. We
obtained the normalized times in Table \ref{table:lebesgue} (the time for
the Moore library was taken as the unit.) This table indicates that
for the arithmetic operations involved in the evaluation of the
Lebesgue function \wref{lebesgue} the Moore library is more
efficient that the Boost, Filib and libieeep1788 libraries.
The difference is slight between Moore and boost (10\%),
more relevant between Moore and Filib (about 300\%) and
very significant between Moore and IeeeP1788 (about 25000\%).

\begin{figure}[!t]
\centering
\includegraphics[height=6.0cm, width=9.0cm]{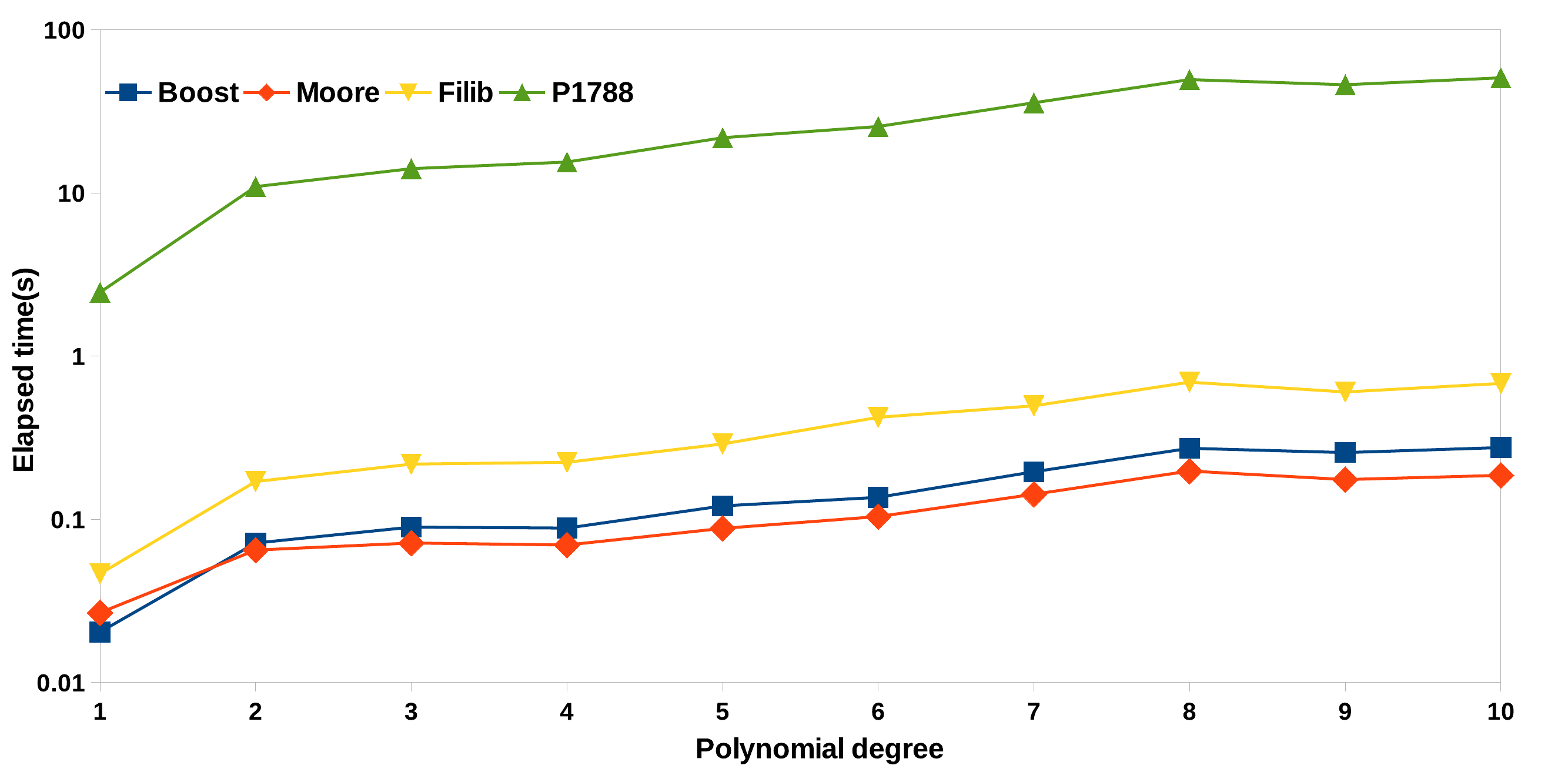}
\caption{Times for Newton's method with polynomials, in log scale.}
\label{fig:newton}
\end{figure}

In our second experiment we considered computation of the roots of
functions which use only arithmetic operations, like the
Lebesgue function in Equation \wref{lebesgue} and its derivatives
with respect to its parameters.
The data for this experiment was generated with
an interval implementation of Newton's method
which can use any one of the four libraries mentioned above.
We compared the times for the solution of random polynomial
equations, with the polynomials and their derivatives
evaluated by Horner's method. We obtained the times
in Figure \ref{fig:newton}, which corroborate the
data in Table \ref{table:lebesgue}.

Our first two experiments show that the Moore library is competitive
for arithmetic operations, but they tell only part of the history about
the relative efficiency of the four libraries considered. In order to
have a more balanced comparison, in our third and last experiment
we compared the times that the four libraries mentioned above take to
evaluate of the elementary functions (sin, cos, exp, etc.)
using the IEEE 754 double precision arithmetic.
The results of this experiment are summarized in Table \ref{table:elem} below,
which shows that in this case the Filib library is far more efficient
that the other three alternatives.

\begin{table}[!t]
\caption{Time for $10^6$ evaluations of the elementary functions with random intervals.}
\label{table:elem}
\centering
\setlength{\tabcolsep}{0.45em}
\begin{tabular}{|l|r|r|r|r|}
\hline
\texttt{Function} & \texttt{Moore} & \texttt{Filib} & \texttt{Boost} & \texttt{P1788}\\
\hline
 $\sin$         & 9.36  &  0.07 &  3.75 &  23.00 \\
 $\cos$         & 8.69  &  0.07 &  3.56 &  21.74 \\
 $\tan$         & 10.86 &  0.06 &  3.94 &  10.44 \\
 \texttt{asin}  &  0.06 &  0.02 &  0.06 &  0.10 \\
 \texttt{acos}  &  0.06 &  0.02 &  0.06 &  0.10 \\
 \texttt{atan}  & 13.24 &  0.06 & 12.34 &  13.29 \\
 $\exp$         &  5.55 &  0.07 &  5.34 &  6.23 \\
 $\log$         &  5.96 &  0.06 &  5.69 &  6.49 \\
\hline
\end{tabular}
\end{table}

Finally, we would like to say that we tried to
be fair with all libraries in the
comparisons presented in this section. To the
best of our knowledge, we used the faster options
for each library. For instance, we used the
boost library on its unprotected mode, which
does not change rounding modes in order to
evaluate arithmetical expressions. The code
was compiled with gcc 6.2.0 with flag
-O3 and {\verb NDEBUG }  defined (the flag
-frounding-math should also be used when compiling
the Moore library.)

\section{Limitations}
The Moore library was designed and implemented using
a novel feature of the C++ language called
concepts \cite{Concepts:cpts}, and it pays the
price for using the bleeding edge of this technology.
The main limitations in the library are due
to the current state of concepts in C++. For instance,
only the latest versions of the gcc compiler
support concepts, and today our
library cannot be used with other compilers.
Concepts are not formally part
of C++ yet, and it will take a few years
for them to reach their final form and become
part of the C++ standard.

Additionally, several decisions regarding
the library were made in order
to get around bugs in gcc's implementations of
concepts and in the supporting libraries we use,
and in order to reduce the compilation
time. Our code would certainly be cleaner if
we did not care about these practical issues,
but without the compromises we took using the
library would be more painful.

Another limitation is the need to
guard the code by constructing an object of type {\small \verb RaiiRounding }.
In other words, the code must look like this
{
\small
\begin{verbatim}
RaiiRounding r;
code using the Moore library
\end{verbatim}
}
\noindent
A similar requirement is made by the most efficient
rounding policy for the boost library, but that
library allows users to choose other policies for rounding,
although the resulting code is less efficient.
Things are different with the Moore library:
as the buyers of Henry Ford's cars in the 1920s, our
users can choose any rounding mode as they want,
so long as it is upwards. Users wanting
to mix code from the Moore library with
code requiring rounding to nearest will
need to resort to kludges like this one:
{\small
\begin{verbatim}
{
Moore::RaiiRounding r;
do some interval operations
}
back to rounding to nearest
{
Moore::RaiiRounding r;
do more interval operations
}
\end{verbatim}
}
\noindent

The experiments with elementary functions
in Section \ref{sec:experiments} show clearly that their
implementation in the Moore
library for IEEE double precision must be improved. In this
specific case, we plan to replace the MPFR library by
our own implementation of the elementary functions, but
this replacement will take time, because there are several
delicate issues to be considered. Hopefully, we will
provide optimized elementary functions for IEEE double
precision in the next version of the Moore library,
which we expect to release in 2018.

Finally, the current version of the Moore library
is the first one and it is quite likely that it contains bugs,
although we did a significant effort to test it.
Its design can certainly be improved,
but we hope that it will evolve with time,
better versions of C++ concepts and
constructive criticism from users.

\appendices
\section*{Acknowledgment}
We would like to thank Tiago Montanher for discussions about this
article and for implemeting the Newton's method used to
generate the data for Fig.\ref{fig:newton}, and for obtaining the
data for Table \ref{table:elem}.
\bibliographystyle{IEEEtran}
\bibliography{IEEEabrv,moore}

\begin{thebibliography}{10}
\providecommand{\url}[1]{#1}
\csname url@samestyle\endcsname
\providecommand{\newblock}{\relax}
\providecommand{\bibinfo}[2]{#2}
\providecommand{\BIBentrySTDinterwordspacing}{\spaceskip=0pt\relax}
\providecommand{\BIBentryALTinterwordstretchfactor}{4}
\providecommand{\BIBentryALTinterwordspacing}{\spaceskip=\fontdimen2\font plus
\BIBentryALTinterwordstretchfactor\fontdimen3\font minus
  \fontdimen4\font\relax}
\providecommand{\BIBforeignlanguage}[2]{{%
\expandafter\ifx\csname l@#1\endcsname\relax
\typeout{** WARNING: IEEEtran.bst: No hyphenation pattern has been}%
\typeout{** loaded for the language `#1'. Using the pattern for}%
\typeout{** the default language instead.}%
\else
\language=\csname l@#1\endcsname
\fi
#2}}
\providecommand{\BIBdecl}{\relax}
\BIBdecl

\bibitem{Boost:boost}
H.~Br{\"o}nnimann, G.~Melquiond, and S.~Pion, ``The design of the boost
  interval arithmetic library,'' \emph{Theoretical Computer Science}, vol. 351,
  no.~1, pp. 111--118, 2006.

\bibitem{CXSC:cxsc}
W.~Hofschuster and W.~Kr{\"a}mer, ``{C-XSC} 2.0: A {C++} library for extended
  scientific computing,'' \emph{Lecture Notes in Computer Science}, vol. 2991,
  pp. 15--35, 2004.

\bibitem{Filib:filib}
M.~Lerch, G.~Tischler, and J.~W. von Gudenberg, ``{filib++}, a fast interval
  library supporting containment computations,'' \emph{ACM Trans. Math.
  Software}, vol.~32, no.~2, pp. 299--324, 2006.

\bibitem{Gaol:gaol}
\BIBentryALTinterwordspacing
F.~Goualard, ``Gaol: Not just another interval library,'' last accessed
  September 21, 2016. [Online]. Available:
  \url{http://www.sourceforge.net/projects/gaol/}
\BIBentrySTDinterwordspacing

\bibitem{Gnu:gnu}
\BIBentryALTinterwordspacing
O.~Heimlich, ``{GNU} octave interval package,'' last accessed September 21,
  2016. [Online]. Available: \url{http://octave.sourceforge.net/interval/}
\BIBentrySTDinterwordspacing

\bibitem{JInterval:jinterval}
D.~Nadezhin and S.~Zhilin, ``J{I}nterval library: principles, development, and
  perspectives,'' \emph{Reliable Computing}, vol.~19, pp. 229--247, 2014.

\bibitem{Lambov:lambov}
B.~Lambov, ``Interval arithmetic using {SSE-2},'' in \emph{Reliable
  Implementation of Real Number Algorithms: Theory and Practice}, ser. Lecture
  Notes in Computer Science, W.~L. P.~Hertling, C. M.~Hoffmann and N.~Revol,
  Eds., vol. 5045.\hskip 1em plus 0.5em minus 0.4em\relax Berlin, Germany:
  Springer, 2008, pp. 102--113.

\bibitem{MPFI:mpfi}
F.~Rouillier and N.~Revol, ``Motivations for an arbitrary precision interval
  arithmetic and the mpfi library,'' \emph{Reliable Computing}, vol.~11, pp.
  275--290, 2005.

\bibitem{NehmeierA:nehmeierA}
M.~Nehmeier, ``libieeep1788: A {C++} implementation of the {IEEE} interval
  standard {P}1788,'' in \emph{Norbert Wiener in the 21st Century {(21CW)}}.

\bibitem{Profil:profil}
\BIBentryALTinterwordspacing
O.~Knueppel, ``{PROFIL/BIAS} - a fast interval library,'' \emph{Computing},
  vol.~53, no. 3--4, pp. 277--287, 1994, last accessed September 21, 2016.
  [Online]. Available: \url{http://www.ti3.tu-harburg.de/knueppel/profil/}
\BIBentrySTDinterwordspacing

\bibitem{Sun:sun}
\BIBentryALTinterwordspacing
\emph{{C++} Interval Arithmetic Programming Reference}, {MC68175/D}, Sun
  Microsystems, 1996, last accessed September 21, 2016. [Online]. Available:
  \url{http://docs.sun.com/app/docs/doc/819-3696-10}
\BIBentrySTDinterwordspacing

\bibitem{P17881:full}
\BIBentryALTinterwordspacing
\emph{1788-2015 IEEE Standard for Interval Arithmetic}, IEEE Std., 2015, last
  accessed September 21, 2016. [Online]. Available:
  \url{https://standards.ieee.org/findstds/standard/1788-2015.html}
\BIBentrySTDinterwordspacing

\bibitem{P17881:p17881}
\BIBentryALTinterwordspacing
\emph{P1788.1-Standard for Interval Arithmetic (simplified)}, IEEE Std., 2015,
  last accessed September 21, 2016. [Online]. Available:
  \url{https://standards.ieee.org/develop/project/1788.1.html}
\BIBentrySTDinterwordspacing

\bibitem{MPFR:mpfr}
L.~Fousse, G.~Hanrot, V.~Lef{\'e}vre, P.~P{\'e}lissier, and P.~Zimmermann,
  ``{MPFR:} a multiple-precision binary floating-point library with correct
  rounding,'' \emph{ACM Trans. Math. Softw}, vol.~33, no.~2, 2007, article 13.

\bibitem{Boost:Bst}
\BIBentryALTinterwordspacing
H.~Br{\"o}nnimann, G.~Melquiond, and S.~Pion, ``The boost interval arithmetic
  library,'' last accessed September 21, 2016. [Online]. Available:
  \url{https://hal.inria.fr/inria-00348711/file/rnc.pdf}
\BIBentrySTDinterwordspacing

\bibitem{MascA}
W.~F. Mascarenhas, ``The stability of barycentric interpolation at the
  chebyshev points of the second kind,'' \emph{Numer. Math.}, vol. 128, pp.
  265--300, 2014.

\bibitem{MascB}
W.~F. Mascarenhas and A.~P. de~Camargo, ``The effects of rounding errors in the
  nodes on barycentric interpolation,'' \emph{Numer. Math.}, 2016,
  doi:10.1007/s00211-016-0798-x.

\bibitem{MascC}
A.~P. de~Camargo and W.~F. Mascarenhas, ``The stability of extended
  floater--hormann interpolants,'' \emph{Numer. Math.}, 2016,
  doi:10.1007/s00211-016-0840-z.

\bibitem{Concepts:cpts}
\BIBentryALTinterwordspacing
W.~Introduction~to C++~concepts, last accessed September 21, 2016. [Online].
  Available: \url{https://en.wikipedia.org/wiki/Concepts\_(C\%2B\%2B)}
\BIBentrySTDinterwordspacing

\end{thebibliography}

\end{document}